\documentclass[runningheads]{svmult}

\usepackage{makeidx}   % allows index generation
\usepackage{graphicx}  % standard LaTeX graphics tool
\usepackage{subeqnar}  % subnumbers individual equations
\usepackage{multicol}  % used for the two-column index
\usepackage{physprbb}  % modified textarea for proceedings,
\makeindex             % used for the subject index
\usepackage{natbib}
\newcommand\M{\ensuremath{\mathcal{M}}}%
\newcommand\Msun{\ensuremath{\M_\odot}}%
\newcommand\Lsun{\ensuremath{L_\odot}}%
\newcommand\Mlim{\ensuremath{\M_{\mathrm{lim}}}}%
\newcommand\MLK{\ensuremath{\M/L_K}}%
\newcommand\Vmax{\ensuremath{V_{\mathrm{max}}}}%

\begin{document}
\title*{The Mass Function of Field Galaxies at $0.4 < z < 1.2$ as Derived
from the MUNICS $K$-Selected Sample}
\toctitle{The Mass Function of Field Galaxies at $0.4 < z < 1.2$ as Derived
from the MUNICS $K$-Selected Sample}
% allows explicit linebreak for the table of content
%
%
\titlerunning{The Mass Function at $0.4 < z < 1.2$}
% allows abbreviation of title, if the full title is too long
% to fit in the running head
%
\author{Niv Drory\inst{1}
\and Ralf Bender\inst{1}
\and Jan Snigula\inst{1}
\and Georg Feulner\inst{1}
\and Ulrich Hopp\inst{1}
\and Claudia Maraston\inst{1}
\and Gary J.\ Hill\inst{2}
\and Claudia Mendes de Oliveira\inst{3}}
\authorrunning{Niv Drory et al.}
% if there are more than two authors,
% please abbreviate author list for running head
%
%
\institute{Universit\"ats-Sternwarte M\"unchen, Scheinerstr. 1,D-81679 M\"unchen, Germany
\and University of Texas at Austin, Austin, Texas 78712, USA
\and Instituto Astron\^omico e Geof\'{\i}sico, Av Miguel St\'efano
  4200, 04301-904, S\~ao Paulo, Brazil}

\maketitle              % typesets the title of the contribution

\begin{abstract}
  We derive the number density evolution of massive field galaxies in
  the redshift range $0.4 < z < 1.2$ using the $K$-band selected field
  galaxy sample from the Munich Near-IR Cluster Survey (MUNICS).  We
  rely on spectroscopically calibrated photometric redshifts to
  determine distances and absolute magnitudes in the rest-frame
  $K$-band. To assign mass-to-light ratios, we use two different
  approaches.  First, we use an approach which maximizes the stellar
  mass for any $K$-band luminosity at any redshift. We take the
  mass-to-light ratio, \MLK , of a Simple Stellar Population (SSP)
  which is as old as the universe at the galaxy's redshift as a likely
  upper limit.  Second, we assign each galaxy a mass-to-light ratio by
  fitting the galaxy's colours against a grid of composite stellar
  population models and taking their \MLK.  We compute the number
  density of galaxies more massive than $2\times 10^{10} h^{-2}
  \Msun$, $5\times 10^{10} h^{-2} \Msun$, and $1\times 10^{11} h^{-2}
  \Msun$, finding that the integrated stellar mass function is roughly
  constant for the lowest mass limit and that it decreases with
  redshift by a factor of $\sim 3$ and by a factor of $\sim 6$ for the
  two higher mass limits, respectively. This finding is in qualitative
  agreement with models of hierarchical galaxy formation, which
  predict that the number density of $\sim M^{*}$ objects is fairly
  constant while it decreases faster for more massive systems over the
  redshift range our data probe.
\end{abstract}

\section{Introduction}
The traditional observables used to characterise galaxies are
unsuitable for studying the assembly history of galaxies, one of the
most fundamental predictions of CDM models, since these observables
may be transient. The best observable for this aim is, in principle,
total mass, which is on the other hand very hard to measure. It has
been argued that the best available surrogate accessible to direct
observation is the near-IR $K$-band luminosity of a galaxy which
reflects the mass of the underlying stellar population and is least
sensitive to bursts of star formation and dust extinction
\citep{RR93,KC98a,BE00}. The main uncertainty involved in the
conversion of $K$-band light to mass is due to the age of the
population, amounting to only a factor of two in mass uncertainty for
populations older than $\sim 3$~Gyr.

The galaxy sample used here is a subsample of the MUNICS survey,
selected for best photometric homogeneity, good seeing, and similar
depth.  Furthermore, in each of the remaining survey patches, areas
close to the image borders in any passband, areas around bright stars,
and regions suffering from blooming are excluded. The subsample covers
0.27 square degrees in $V$ (23.5), $R$ (23.5), $I$ (22.5), $J$ (21.5),
and $K$ (19.5); the magnitudes are in the Vega system and refer to
50\% completeness for point sources.

The final catalog covers an area of 997.7 square arc minutes and
contains 5132 galaxies. The fields included in this analysis are S2,
S3f5--8, S5, S6, and S7f5--8. See Table 1 in \cite{MUNICS1} for
nomenclature and further information on the survey fields.

The distances to the galaxies are derived using spectroscopically
calibrated photometric redshifts. A comparison of spectroscopic and
photometric redshifts is shown in \citet{MUNICS3}.

We derive stellar masses by converting rest-frame $K$-band
luminosities to mass using two different approaches to model the
mass-to-light ratios of the galaxies.  We discuss the resulting
integrated stellar mass functions at different mass limits and their
evolution with redshift.

We assume $\Omega_M = 0.3$, $\Omega_{\Lambda} = 0.7$ throughout this
work. We write Hubble's Constant as $H_0 = 100\ h\ \mathrm{km\ 
  s^{-1}\ Mpc^{-1}}$, using $h = 0.65$ unless the quantities in
question can be written in a form explicitly depending on $h$.

\section{The maximum PLE model}

The integrated stellar mass function $n(\M>\Mlim)$, the comoving
number density of objects having stellar mass exceeding \Mlim, is
computed using the \Vmax\ formalism as desctibed in \cite{MUNICS3}.

To compute the stellar mass of a galaxy, we first use an approach
which maximises the stellar mass for any $K$-band luminosity at any
redshift. 

Noting that \MLK\ is a monotonically rising function of age for Simple
Stellar Populations (SSPs), we find that the likely upper limit for
\MLK\ is the mass-to-light ratio of a SSP which is as old as the
universe at the galaxy's redshift. This is the most extreme case of
passive luminosity evolution (PLE) one can adopt. It corresponds to a
situation where all massive galaxies would be of either elliptical,
S0, or Sa type.

We take the mass-to-light ratios from the SSP models published by
\citet{Maraston98}, using a Salpeter IMF. Similar dependencies on age
are obtained from the models of \citet{Worth94} and \citet{BC93}
although the absolute values of \MLK\ vary somewhat, partly due to
differences in the models themselves but mostly due to the way stellar
remnants are treated by the different authors.

The resulting integrated mass functions for $\Mlim = 2\times 10^{10}
h^{-2} \Msun$, $\Mlim = 5\times 10^{10} h^{-2} \Msun$, and $\Mlim =
1\times 10^{11} h^{-2} \Msun$ are shown in Fig.~\ref{f:integ_mf3}
along with the integrated luminosity functions for comparison. 

The mean values of \MLK\ in the maximum PLE model in the four redshift
bins are $0.99$, $0.83$, $0.73$, and $0.65$, as computed from the
look-back time in our cosmology. With these mean values the mass
limits correspond to absolute $K$-band magnitudes of $-22.43$,
$-22.63$, $-22.77$, and $-22.90$, respectively, for $\Mlim = 2\times
10^{10} h^{-2} \Msun$.  For $\Mlim = 5\times 10^{10} h^{-2} \Msun$ the
numbers are $-23.42$, $-23.62$, $-23.76$, and $-23.89$. Finally, for
$\Mlim = 1\times 10^{11} h^{-2} \Msun$ we have $-24.18$, $-24.38$,
$-24.51$, and $-24.64$ (magnitudes with respect to $h = 1$).

The upper and middle panels of Fig.~\ref{f:integ_mf3} compare the
evolution of the integrated luminosity to the integrated mass.  It is
evident that the number density of {\em luminous} $K$-band selected
galaxies does not evolve significantly (given our uncertainties) to $z
= 1.2$.  However, because of the inevitable evolution of the
mass-to-light ratio with $z$, the number density of {\em massive}
systems does change.  Transforming luminosities into masses with our
maximum PLE scheme yields a roughly constant number density for our
lowest mass limit, $2\times 10^{10} h^{-2} \Msun$, and a decrease of
the number density with redshift by a factor of $\sim 3$ for a mass
limit of $5\times 10^{10} h^{-2} \Msun$, and by a factor $\sim 6$ for
objects more massive than $1\times 10^{11} h^{-2} \Msun$.  As the true
\MLK\ at high redshift will most likely be lower than in our maximum
PLE model, the true number densities are likely to decrease more
rapidly with redshift.

The steepening of the curves with increasing limiting mass in the
maximum PLE models (despite them all having the same mass-to-light
ratios at any given redshift) is due to the invariance of the LF with
redshift and its steepness at the bright end. At increasing limiting
mass, one is moving down the steepening bright end of the LF, so that
the same change in the mass-to-light ratio yields a higher change in
the number density.

To investigate the effect the uncertainties in the photometric
redshifts have on the values of the integrated mass function, we have
performed Monte-Carlo simulations. The errors of the mean values of
the integrated mass function (size of open symbols in
Fig.~\ref{f:integ_mf3}) are derived by repeating the mass function
analysis using subsamples of the template SED library, as deficiencies
in the templates are the main source of concern for the accuracy of
the redshifts during the photometric redshift determination.

\section{The fitted mass-to-light-ratio model}

To obtain a more realistic estimate of \MLK, we used our VRIJK color
information and the photometric redshift to fit the age and SFR of
each galaxy using a grid of composite stellar populations (CSP) with 9
exponential star formation timescales, $\tau$, ranging from 0.1 to
10~Gyr with spectra extracted for 28 ages, $t$, between 0.04~Gyr and
15~Gyr for each $\tau$. The input SSPs for constructing the composite
stellar populations models are taken from \citet{Maraston02}, again
using Salpeter IMF.

\begin{figure}[ht]
  \centering
  \includegraphics[width=0.45\textwidth]{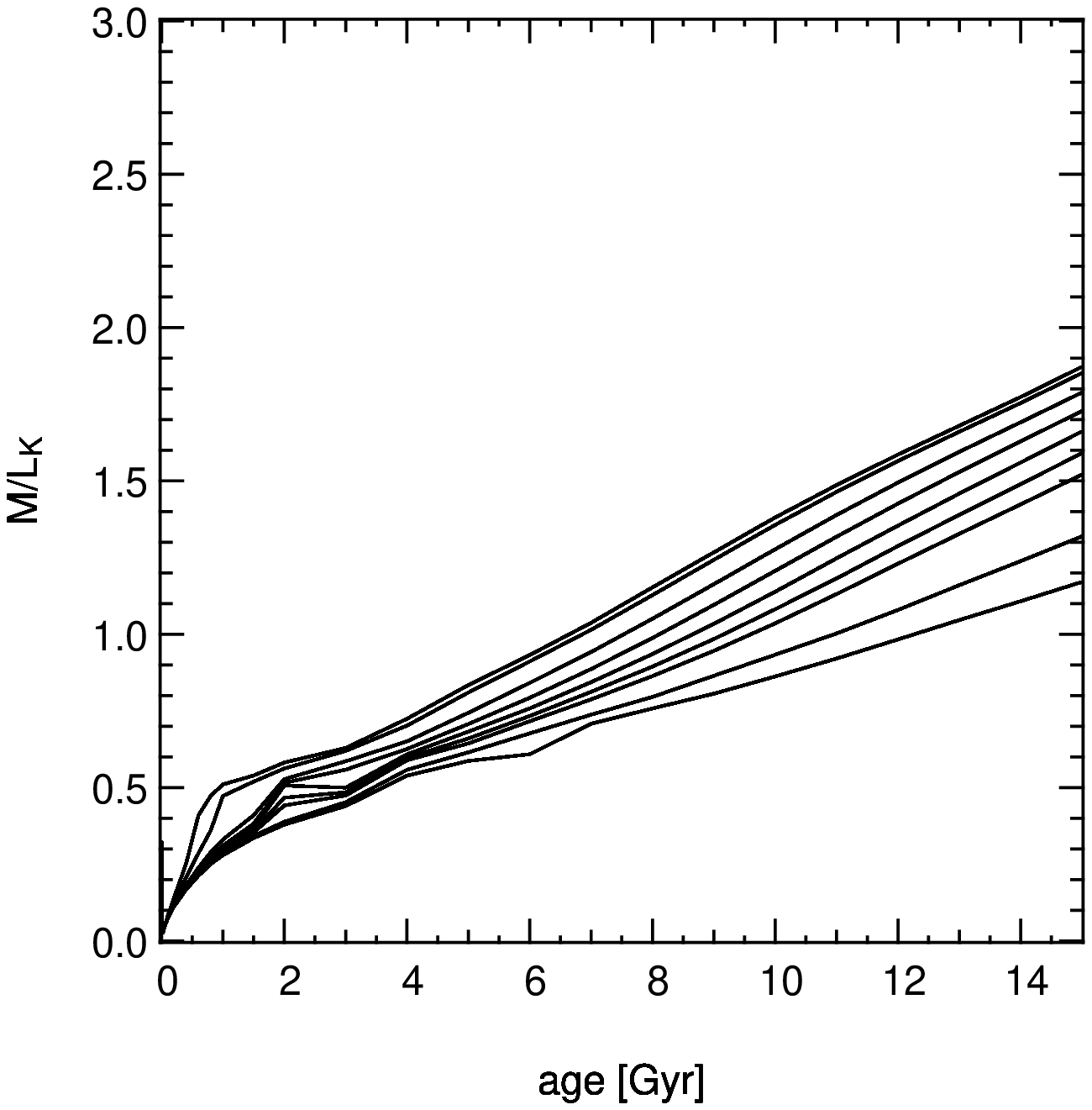}\hfill
  \includegraphics[width=0.45\textwidth]{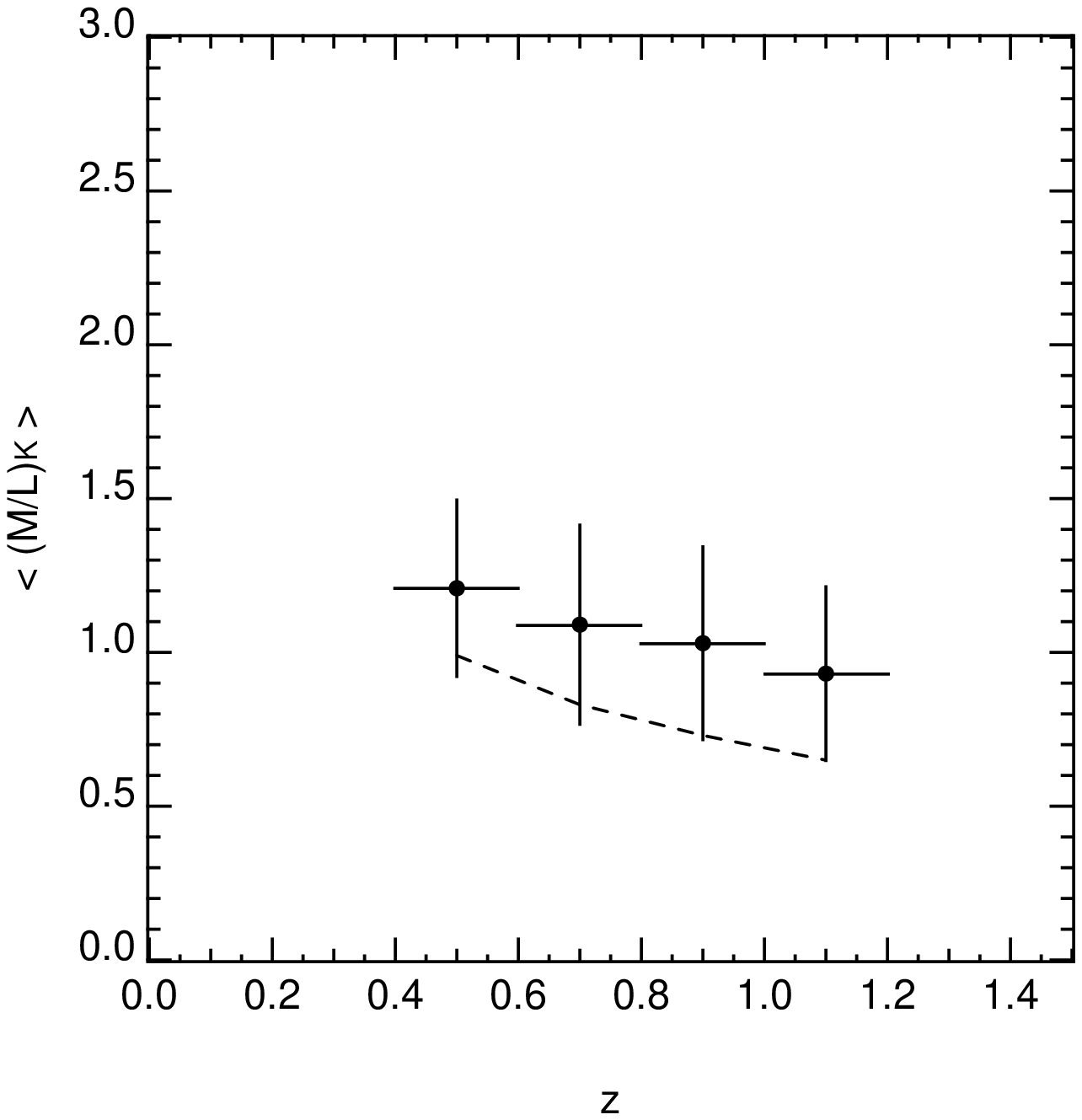}
  \caption{Evolution of the $K$-band mass-to-light ratio, \MLK,
    with age for different star formation histories (left panel; see
    text). The average $K$-band mass-to-light ratio,\MLK, of the
    MUNICS sample as a function of redshift as determined from fitting
    CSP models to the $V,R,I,J,K$ colour data base. The dashed line
    denotes the mass-to-light ration of the maximum PLE model (see
    text; right panel).}
  \label{f:mlk_mean}
\end{figure}

Fig.~\ref{f:mlk_mean} shows the evolution of the $K$-band
mass-to-light ratios as a function of age for each value of $\tau$ in
the grid. Except for the two largest values of $\tau$, the slope at
ages $t > 2$~Gyr is remarkably independent of the actual star
formation timescale. Moreover, the slope of the time evolution of
\MLK\ is the same even for the shortest value of $\tau$, 0.1~Gyr,
which essentially represents an SSP. Let aside normalisation effects,
we therefore may expect a similar result for the mass function as
obtained with the PLE model.

The average $K$-band mass-to-light ratio of the galaxy population
determined by applying this fitting procedure is shown in
Fig.~\ref{f:mlk_mean}. The figure also shows the PLE mass-to-light ratio
as a function of $z$. Apart from the different normalisation, the
evolution with redshift is very similar, a consequence of the
insensitivity of \MLK\ on the star formation history.

\begin{figure}
  \includegraphics[width=\textwidth]{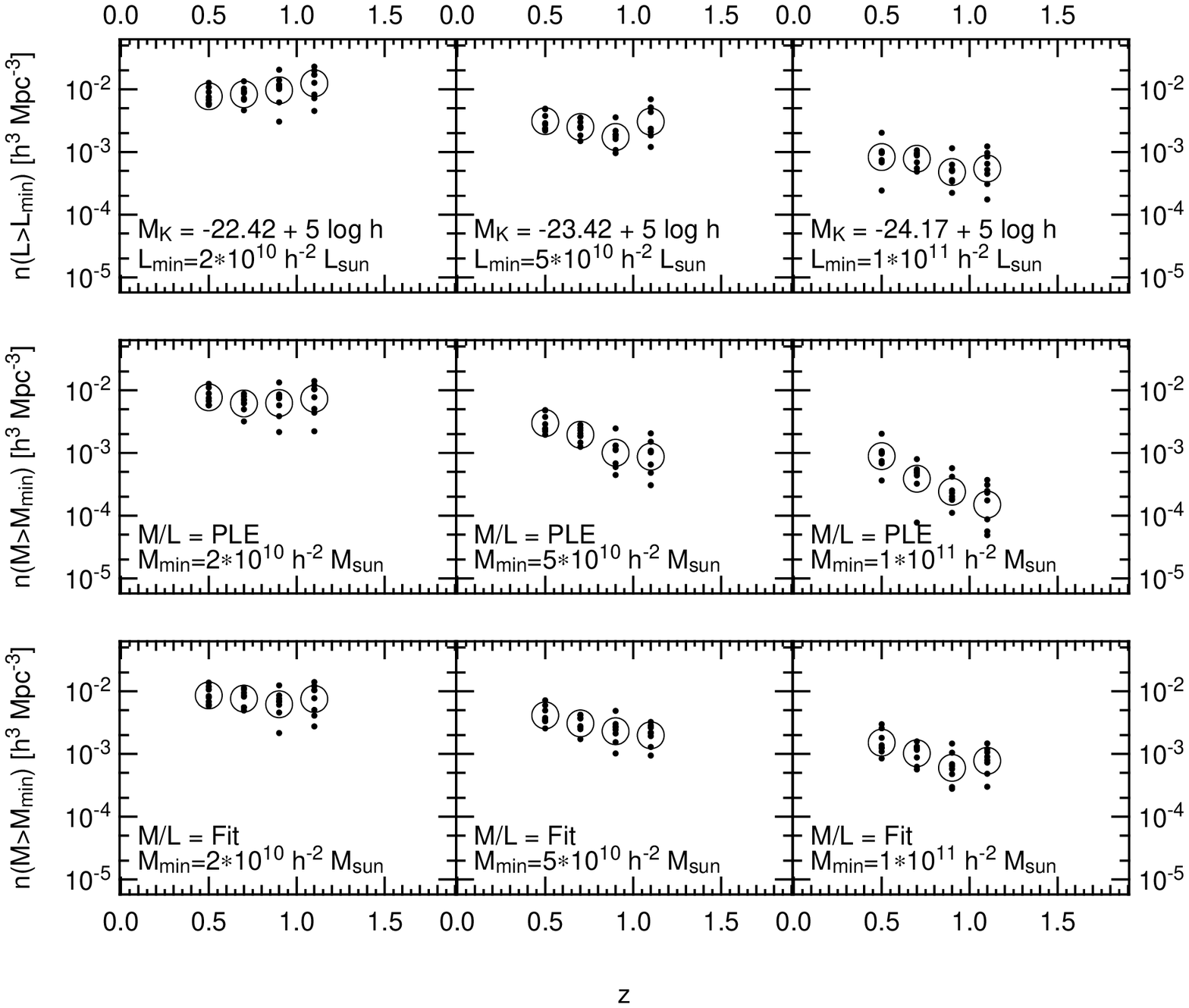}
  \caption[Inegrated mass function using PLE and
  fitted mass-to-light ratios]{Comoving number density of objects
    having rest-frame $K$-band luminosities exceeding $-22.42 + 5\log
    h$ ($2\times 10^{10} h^{-2} \Lsun$), $-23.42 + 5\log h$ ($2\times
    10^{10} h^{-2} \Lsun$), and $-24.17 + 5\log h$ ($2\times 10^{10}
    h^{-2} \Lsun$) (upper panels) and comoving number density of
    objects having stellar masses exceeding $\Mlim = 2\times 10^{10}
    h^{-2} \Msun$, $\Mlim = 5\times 10^{10} h^{-2} \Msun$, and $\Mlim
    = 1\times 10^{11} h^{-2} \Msun$ (integrated stellar mass
    functions; middle panels).  Mass to light ratios are assigned to
    maximise the stellar mass at a given luminosity (see text), and
    thus are likely upper limits. The lower panel shows the integrated
    mass function for the same mass limits, this time individually
    determining \MLK\ for each object by fitting against a grid of CSP
    models (see text). The solid points denote the values measured
    separately in each survey field, the open circles denote the mean
    values over the whole survey area.  The size of the open circles
    is chosen to represent our estimate of the total uncertainty in
    the mean values.}
  \label{f:integ_mf3}
\end{figure}

Finally, the lower panel of Fig.\ref{f:integ_mf3} shows the integrated
mass function for the same mass limits as those applied above, using
the individually fitted \MLK\ values.

The most striking feature of Fig.~\ref{f:integ_mf3} is the similarity of
the maximum PLE and the CSP-fitted curves. Note that there is a
difference in the normalisation of the two, and due to the log scaling
of the figure, the slope appears to be different in the plot.

If star formation played an important role at $z \sim 1$, the presence
of young populations would have pushed \MLK\ down, and therefore the
CSP-fitted curves would be expected to be steeper than the maximum PLE
curve, which assumes no star formation happens at all after $z =
\infty$. 

Nevertheless, the number density of massive systems seems to decline,
with this decline being stronger for more massive systems, and
therefore one is inclined to think that merging does play an important
role. Indeed, \citet{HSTCFRS100} derive a number of 0.6 to 1.8 major
mergers per $L^*$ galaxy since $z \sim 1$ from HST-based pair counts
of galaxies with known redshifts selected from the CFRS. We observe a
decline in the number density by a factor of $> 2$ for somewhat more
massive systems, and almost no significant density evolution at $L^*$.

Therefore, we are inclined to think that if merging is the dominant
factor in increasing the mass of these $K$-selected massive galaxies,
most of the merging has to be dissipationless, involving rather low
star formation activity.

The main uncertainty in these conclusions is still the field to field
variation, in spite of the relatively large area surveyed, followed by
the choice of SED templates used in the photometric redshift code (see
above). The size of the open symbols in Fig.~\ref{f:integ_mf3}
represents our estimate of the total uncertainty of the mean values.
If we assume a Gould IMF \citep{GFB98} instead of a Salpeter IMF, the
evolving \MLK\ curve becomes lower in its normalization as the
mass-to-light ratio becomes smaller due to the reduced number of
low-mass stars. The slope does not change significantly.

The observed density evolution as a function of mass is qualitatively
consistent with the expectation from hierarchical galaxy formation
models. Most rapid evolution is predicted for the number density of
the most massive galaxies while the number density of $L^*$-galaxies
should evolve much less.  E.g.\ \citet{BCFL98} predict that the number
density of galaxies of a stellar mass of $10^{10} h^{-1} \Msun$
decreases by a factor of $\sim 3.1$ over redshift range $0.4 < z <
1.2$ (for the cosmological parameters as used here).  Though this
agreement is encouraging, both more elaborated models and improved
sets of data are required.  The latter can be obtained by larger and
deeper samples, and more realistic estimates of \MLK\ based on
spectroscopic observations of the galaxies..

\subsection*{Acknowledgments}
We would like to thank the Calar Alto staff for their long-standing
support. This work was partly supported by the Deutsche
Forschungsgemeinschaft, grant SFB 375 ``Astroteilchenphysik'' and the
German Federal Ministry of Education and Research (BMBF), grant 05
AV9WM1/2.

%INDEX%%%%%%%%%%%%%%%%%%%%%%%%%%%%%%%%%%%%%%%%%%%%%%%%%%%%%%%%%%%%%%%
% Please check with the editor of your book whether he plans to
% include a "mutual" subject index - if so, please code your entries
% in the standard syntax. For your own purposes you may print your
% "personal" index by using the following commands:
%
%\clearpage
%\addcontentsline{toc}{section}{Index}
%\flushbottom
%\printindex
%%%%%%%%%%%%%%%%%%%%%%%%%%%%%%%%%%%%%%%%%%%%%%%%%%%%%%%%%%%%%%%%%%%%%
\end{document}